\documentclass[conference]{IEEEtran}
\usepackage[pdftex]{graphicx}
\usepackage{multirow}
\usepackage{nicefrac}
\usepackage{url}

\begin{document}
\title{Statistics of Reflection and Transmission in the Strong Overlap Regime of Fully Chaotic Reverberation Chambers}
\author{\IEEEauthorblockN{Ulrich Kuhl, Olivier Legrand, Fabrice Mortessagne, Khalid Oubaha, Martin Richter}
\IEEEauthorblockA{Institute of Physics in Nice, University of C\^{o}te d'Azur, 06100 Nice, France\\
Emails: forename.surname@unice.fr}
}

\IEEEspecialpapernotice{(Invited Paper for the special session 'Stochastic Electromagnetics' of the EMC at EUMCWeek 2017 in N\"{u}rnberg)}
\maketitle

\begin{abstract}
\boldmath
We have experimentally investigated a chaotic reverberation chamber in the regime of strong modal overlap ($1{<}d{<}150$) varying the opening as well as the coupling strength $\kappa$ of the two attached antennas. We find a good agreement with numerical distribution of the reflection $R=|S_{ii}|^2$ and transmission $T=|S_{21}|^2$ obtained via the effective Hamiltonian formalism using random matrix theory. The two parameters entering the numerics, $\kappa$ and the decay rate $\tau$ were determined beforehand experimentally.
Additionally we verified the relation predicted by Schroeder and Kuttruff for acoustic rooms between the averaged frequency spacing of maximal transmission $\langle\delta f_{max}\rangle$ and the decay rate.
\end{abstract}

\IEEEpeerreviewmaketitle

\section{Introduction}
Reverberation chambers play a crucial role in electromagnetic compatibility testing.
For the characterization of reverberation chambers it is assumed that the field inside the cavity is isotropic and the field components follow a bivariate Gaussian distribution.
In the electromagnetic community this so called Hill's hypothesis\cite{hil98} is typically realized when the resonance overlap is large.
It is also realized in chaotic cavities once one is sufficiently far from the lowest eigenfrequency and is related to Berry's ansatz\cite{ber77a}.
The strong overlap regime is then known as Ericson regime\cite{eri63,eri16}
in nuclear physics and as the regime of universal conductance fluctuations in mesoscopic physics\cite{lee85b},
even though there are slight differences in the definition.
This regime has also been known in room acoustics for a long time\cite{sch62a}.
We will here investigate the effects of the modal overlap on the distribution of the reflection and transmission and compare it to predictions using random matrix theory (RMT) numerics as well as verify the prediction by Schroeder and Kuttruff for the mean frequency spacing of transmission maxima to the average decay rate of the system\cite{sch62a,sch62b,kut09}.

\section{Scattering and Random Matrix Theory using the Effective Hamiltonian}

\begin{figure}[!t]
\centering
\includegraphics[width=.69\linewidth]{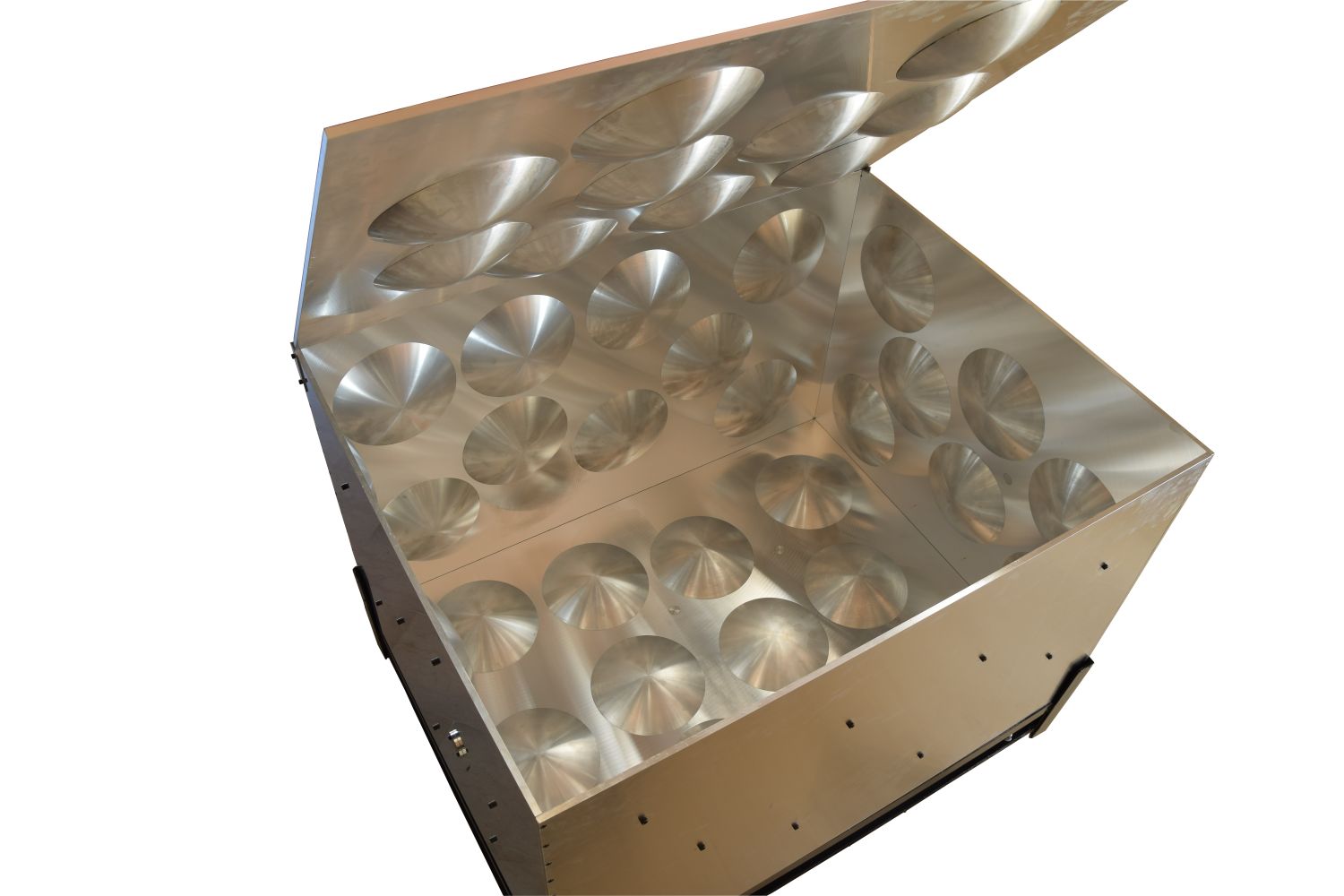}
\caption{\label{fig::photo_crc}
Photograph of the chaotic reverberation chamber (CRC) with length $L$=100\,cm, width $W$=77\,cm and height $H$=62\,cm.
At the walls 54 spherical caps of radius $r_c$=10\,cm and cap height $h_c$=3\,cm are used.
The total internal volume is $V$= 0.451\,m$^3$. Two monopole antennas have been attached to the side walls ranging 1.5\,cm into the CRC.
}
\end{figure}

\begin{figure}[!t]
\centering
\includegraphics[width=.9\linewidth]{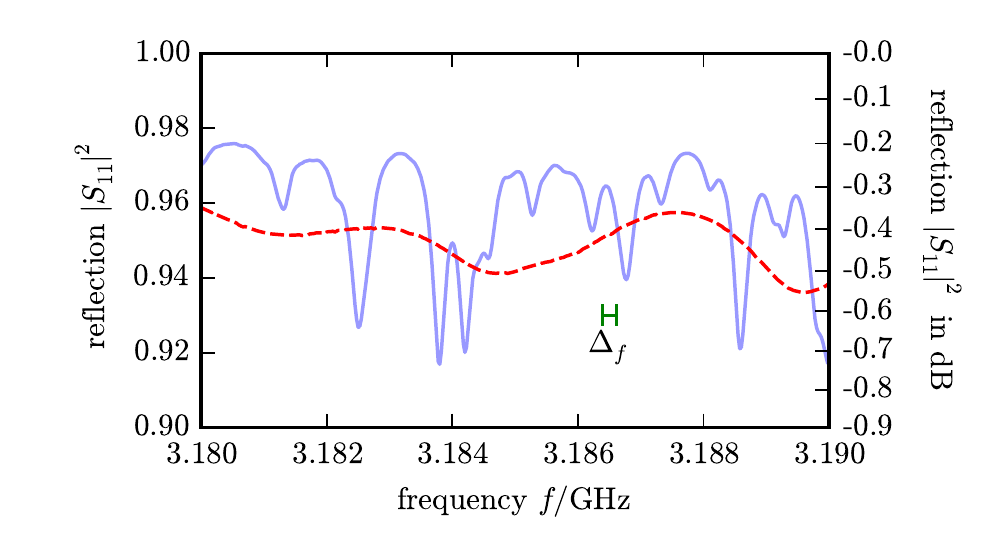}\\[-0.5cm]
\includegraphics[width=.9\linewidth]{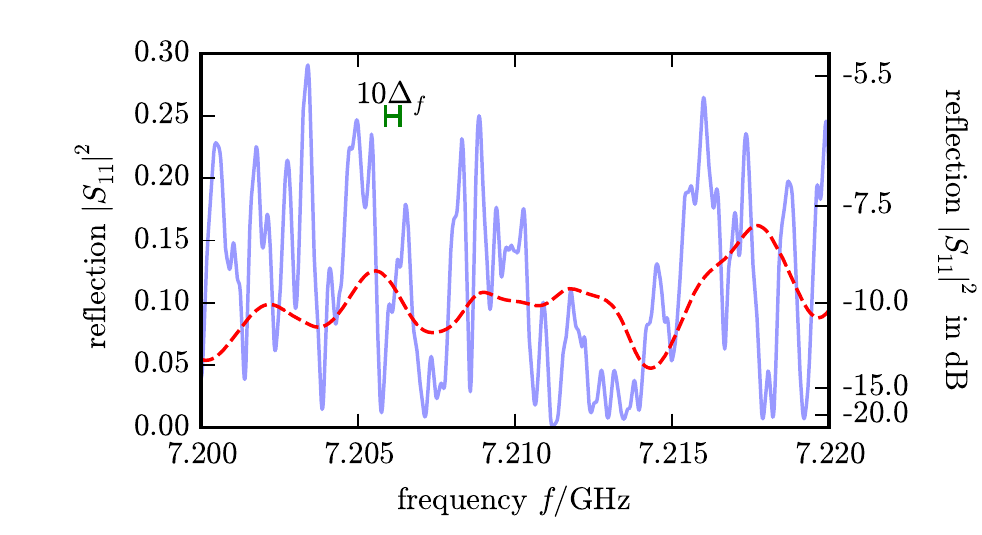}\\[-0.5cm]
\includegraphics[width=.9\linewidth]{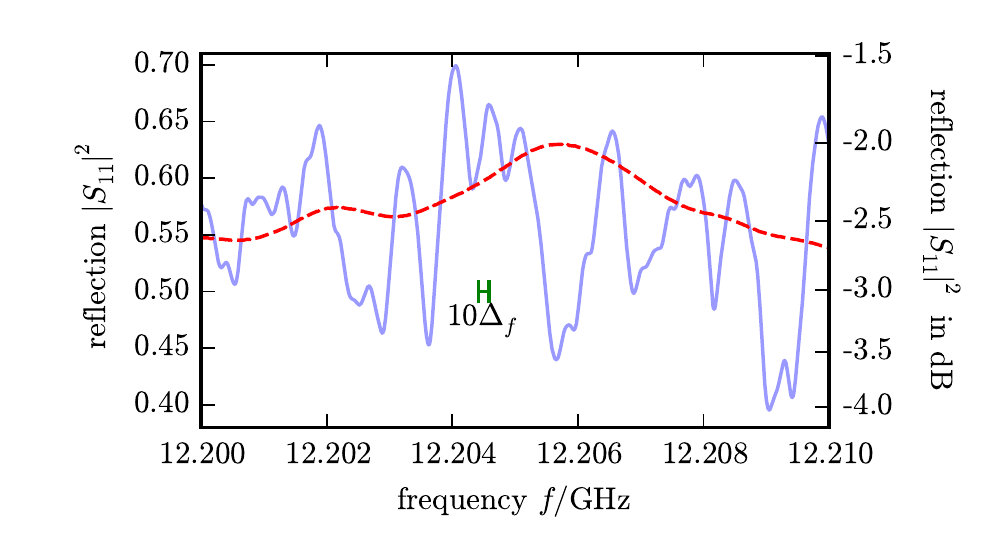}\\[-0.2cm]
\caption{\label{fig::RSpectra}
Measured reflection $R=|S_{11}|^2$ for three different frequency ranges and for two different openings ($h_o$=0\,mm and 24\,mm).
The reflection for the closed CRC is shown in solid light blue ($h_o$=0\,mm), whereas the open system ($h_o$=24\,mm) is shown in dashed red.
The corresponding mean frequency spacing $\Delta_f$ is indicated by the bar.
}
\end{figure}

\begin{figure}[!t]
\centering
\includegraphics[width=.9\linewidth,clip=false]{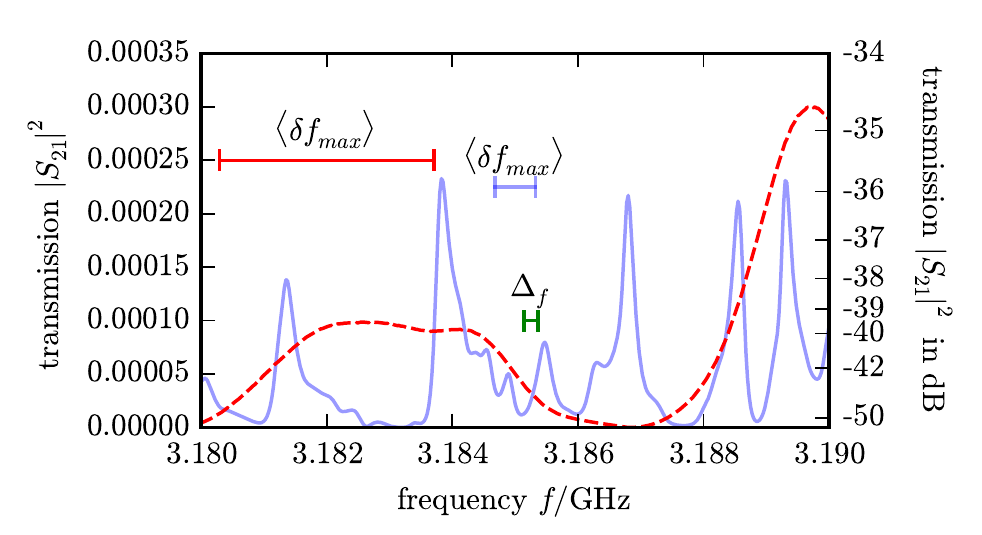}\\[-0.5cm]
\includegraphics[width=.9\linewidth,clip=false]{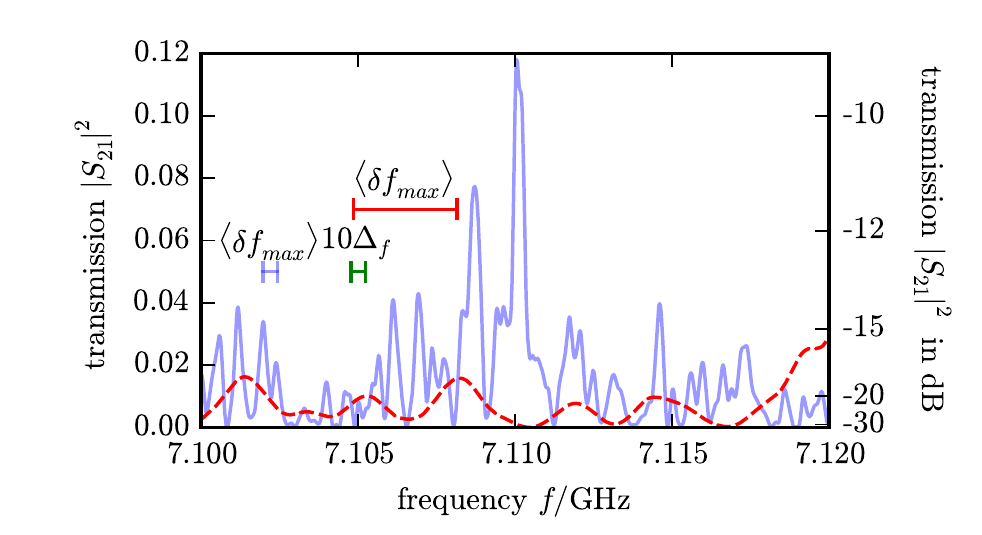}\\[-0.5cm]
\includegraphics[width=.9\linewidth,clip=false]{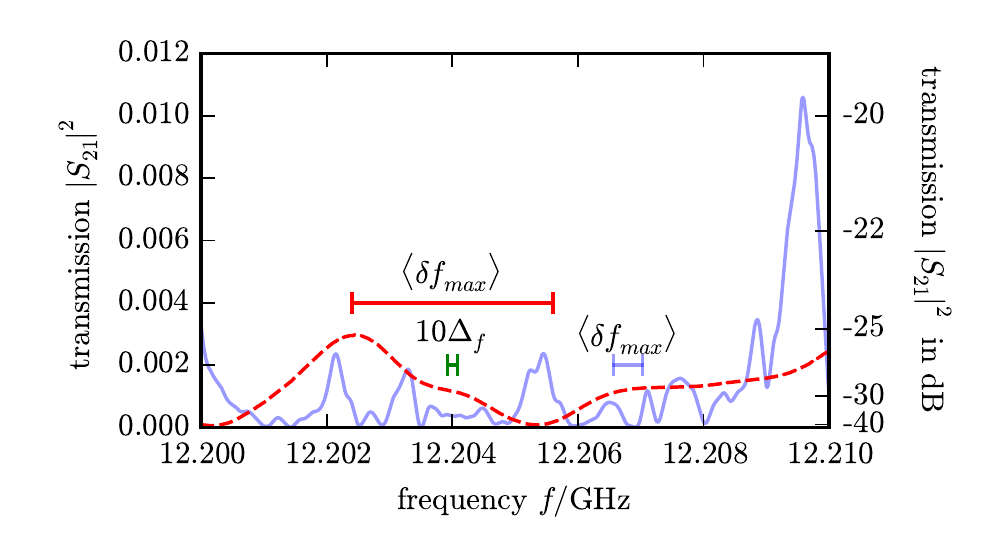}\\[-0.2cm]
\caption{\label{fig::TSpectra}
Measured transmission $T=|S_{21}|^2$ for three different frequency ranges and for two different openings ($h_o$=0\,mm and 24\,mm).
The transmission for the closed CRC is shown in solid light blue ($h_o$=0\,mm), whereas the open system ($h_o$=24\,mm) is shown in dashed red.
The corresponding mean frequency spacing $\Delta_f$ and the mean spacings of the maxima $\langle\delta f_{max}\rangle$ are indicated for the two systems by bars, correspondingly.
}
\end{figure}

To describe scattering of waves, the scattering matrix $S$ is typically the main subject of investigation.
It describes the relation between the incoming waves $\Psi_{in}$, here guided by the microwave cables and injected by two monopole antenna, and the outgoing waves $\Psi_{out}$.
If the system is sufficiently complex (e.g. see fig.~\ref{fig::photo_crc}) the idea of a deterministic prediction or calculation of the scattering amplitudes is not reasonable: instead, statistical predictions become relevant.
In particular, in chaotic reverberation chambers, the statistical requirements of the international standard IEC 61000-4-21 have been shown to be ideally verified even for frequencies close to or below the lowest usable frequency (LUF) \cite{gro15,gro16}.

Thus, instead of describing the details of the system, only the main features of the system are taken into account.
Our chaotic reverberation chamber (CRC) is described by a Hamiltonian $H$ which corresponds to a system with time-reversal invariance and is statistically drawn from the Gaussian Orthogonal Ensemble (GOE)\cite{gro16}.
The scattering is then described by the number $M$ of attached channels, in our case the monopole antennas, and their coupling strength $\kappa$ or their channel transmission $T_a$. Additionally, absorption is present in electromagnetic cavities which we take into account by giving the energy an additional imaginary part.
Thus the scattering matrix elements are given by
\begin{equation}\label{eq:def_S}
 S_{ab}(E) = \delta_{ab} - i\sqrt{\mathrm{Re}(\kappa_a)\mathrm{Re}(\kappa_{b})}\, V_a^{\dag}\frac{1}{E-H_{\mathrm{eff}}}V_{b}\,.
\end{equation}
and the effective Hamiltonian is given by
\begin{equation}\label{eq:Heff}
 H_{\mathrm{eff}} = H - \frac{i}{2}\sum_{c=1}^{M}\kappa_c V_cV_c^{\dag}\,.
\end{equation}
where $H$ is a Hamiltonian described by an $N\times N$ matrix and $V$ an $N\times M$ matrix describing the coupling of the $M$ channels to the system.
\begin{equation}\label{eq:T_a}
T_i=1-|\langle S_{ii}\rangle|^2=4\mathrm{Re}(\kappa) / |1+\kappa|^2
\end{equation}
and thus once the scattering elements $S_{ii}$ are measured the coupling strength $\kappa$ can be determined by
\begin{equation}\label{eq:kappa}
\kappa=\frac{|1-\langle S_{ii}\rangle|^2}{1-|\langle S_{ii}\rangle|^2}= \frac{(2 - T_i) \pm 2\sqrt{1 - T_i}}{T_i},
\end{equation}
where we used always the smaller $\kappa$ value.
The imaginary part of the energy $\langle \Gamma\rangle/2$ can be calculated using the intensity decay time $\tau=\langle \Gamma\rangle^{-1}$ of the Fourier transformed transmission signal given by
\begin{equation}\label{eq:tau}
I(t)=|FT(S_{21})|^2=I_0e^{-t/\tau}.
\end{equation}
For details on the effective Hamiltonian approach, couplings and its relations to experiments see Refs.~\cite{ver85a,guh98,koeb10,kuh13,rot15}.
Have in mind that all formulations presented here are written using the quantum mechanical formalism based on the Schr\"{o}dinger equation and the energy is scaled by the mean level spacing $\Delta_E$.

Predictions can be transferred to other kinds of waves, where mainly one has to care about the mean level spacing.
Even though the electromagnetic fields are vectorial fields, the whole formalism can be used, though some minor redefinitions might be necessary\cite{gro14a,gro15,gro16}.
To fix the values used for the RMT calculations we need the mean frequency spacing in three dimensional electromagnetic cavities, given by
\begin{equation}\label{eq:Delta_f}
\Delta_f=\frac{c^3}{8\pi V f^2},
\end{equation}
where $V$ is the volume of the cavity, $c$ the speed of light and $f$ the mean of the measured frequency window.
The quality factor can be retrieved by the decay rate
\begin{equation}\label{eq:Q}
Q=2\pi f / \langle \Gamma \rangle=2\pi\tau f
\end{equation}
thus leading to the averaged modal overlap of
\begin{equation}\label{eq:d}
d=\frac{\langle\Gamma\rangle}{2\pi \Delta_f}=\frac{8\pi V}{Q}\left(\frac{f}{c}\right)^3.
\end{equation}
Now all parameters entering the calculations are fixed leaving no free parameter in the description.

\section{Experimental setup}

The reverberation chamber used here is made of Aluminum. It is a rectangular cavity, where 54 spherical caps were placed on the walls (details see fig.~\ref{fig::photo_crc}). Two monopole antennas are inserted via holes with 2\,mm radius through the 8\,mm thick side walls, which have a length inside the cavity of 10\,mm. The total volume is $V$= 0.451\,m$^3$.
We have performed measurements in four different frequency regimes (3-3.5, 7-7.5, 7.5-8, and 12-12.25\,GHz) and each measurement contained 20001 frequency points.
These frequency ranges were measured with six different openings ($h_o$=0, 4, 10, 14, 24, 30\,mm), where the ceiling was lifted by $h_o$.

\begin{table}[!t]
\caption{\label{tab:InfoMeas}
Information on the different measurements performed in the CRC.
Table of the mean resonance frequency spacing $\Delta_f$, the decay rate $\tau$, the quality factor $Q$, the mean modal overlap $d$, the antenna couplings $T_a$ and $T_b$ and the average spacing of transmission maxima $\langle\delta f_{max}\rangle$.
}
\centering

\begin{tabular}{|c|r|r|r|r|r|r|r|r|r|}
\hline
$\frac{f}{\textrm{GHz}}$ & $\frac{\Delta_f}{\textrm{kHz}}$
& $\frac{h_o}{\textrm{mm}}$
& $\frac{\tau}{\textrm{ns}}$ &
$\frac{Q}{10^{3}}$ & $d$ & $T_\mathrm{a}$ & $T_\mathrm{b}$ &
                                                      $\frac{\delta\!f_\mathrm{max}}{\textrm{kHz}}$\\
\hline
 \multirow{6}{.5cm}{3.0 \linebreak\ -\linebreak 3.5}&
\multirow{6}{*}{225}
   &  0 & 663 & 13.5 &   1.1 & 0.047 & 0.045 &   646\\
&  &  4 & 467 &  9.5 &   1.5 & 0.047 & 0.046 &   816\\
&  & 10 & 176 &  3.6 &   4.0 & 0.047 & 0.045 &  1936\\
&  & 14 & 135 &  2.8 &   5.2 & 0.047 & 0.045 &  2397\\
&  & 24 &  84 &  1.7 &   8.4 & 0.047 & 0.046 &  3420\\
&  & 30 &  84 &  1.7 &   8.4 & 0.047 & 0.046 &  3997\\
\hline
\multirow{6}{.5cm}{7.0 \linebreak\ -\linebreak 7.5}&
\multirow{6}{*}{ 45}
   &  0 & 628 & 28.6 &   5.6 & 0.883 & 0.864 &   453\\
&  &  4 & 396 & 18.1 &   8.9 & 0.883 & 0.863 &   725\\
&  & 10 & 152 &  6.9 &  23.2 & 0.882 & 0.864 &  1922\\
&  & 14 & 136 &  6.2 &  25.8 & 0.882 & 0.863 &  2163\\
&  & 24 &  89 &  4.1 &  39.4 & 0.883 & 0.864 &  3310\\
&  & 30 &  66 &  3.0 &  53.3 & 0.882 & 0.866 &  4593\\
\hline
\multirow{6}{.5cm}{7.5 \linebreak\ -\linebreak 8.0}&
\multirow{6}{*}{ 40}
   &  0 & 640 & 31.2 &   6.3 & 0.730 & 0.708 &   481\\
&  &  4 & 394 & 19.2 &  10.2 & 0.731 & 0.709 &   743\\
&  & 10 & 159 &  7.7 &  25.3 & 0.733 & 0.707 &  1927\\
&  & 14 & 139 &  6.8 &  28.8 & 0.734 & 0.707 &  2042\\
&  & 24 &  79 &  3.9 &  50.6 & 0.730 & 0.709 &  3596\\
&  & 30 &  68 &  3.3 &  59.6 & 0.732 & 0.706 &  4297\\
\hline
\multirow{6}{.5cm}{12.0 \linebreak\ -\linebreak 12.25}&
\multirow{6}{*}{ 16}
   &  0 & 663 & 50.5 &  14.8 & 0.434 & 0.448 &   462\\
&  &  4 & 294 & 22.4 &  33.5 & 0.430 & 0.449 &   998\\
&  & 10 & 188 & 14.4 &  52.3 & 0.433 & 0.447 &  1572\\
&  & 14 & 148 & 11.3 &  66.4 & 0.431 & 0.447 &  1967\\
&  & 24 &  90 &  6.9 & 109.0 & 0.432 & 0.446 &  3203\\
&  & 30 &  65 &  5.0 & 150.4 & 0.427 & 0.449 &  4199\\
\hline
\end{tabular}
\end{table}

Figures~\ref{fig::RSpectra} and \ref{fig::TSpectra} present the measured reflection and transmission spectra for the closed ($h_o$=0\,mm) and one open system ($h_o$=24\,mm).
The corresponding mean level spacing is indicated by small bars and in the transmission figures additionally the mean spacing between transmission maxima $\langle\delta f_{max}\rangle$ is indicated.
The extracted mean resonance spacing $\Delta_f$, the decay rate $\tau$ obtained via a fit of I(t), the quality factor $Q$, and the two antenna transmissions  $T_a$ and $T_b$ are given in table~\ref{tab:InfoMeas}.
Thus all parameters necessary to calculate the statistical predictions from the effective Hamiltonian are extracted directly from the experiment.
We would like to note that a range from 1 to 150 is spanned for the modal overlap $d$.
We found that the normalized $S$-matrix elements $|S_{ij}-\langle S_{ij}\rangle|/\langle|S_{ij}-\langle S_{ij}\rangle|\rangle$ followed the predicted bivariate distribution for large overlaps $d$\cite{die10a}.

\section{Results}

\subsection{Distribution of Reflection and Transmission}
\begin{figure}[!t]
\centering
\includegraphics[width=.85\linewidth]{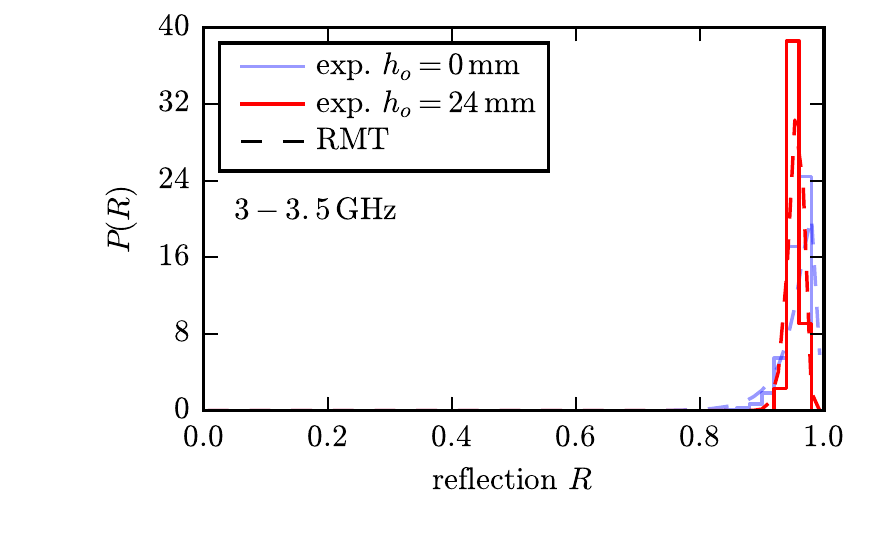}\\[-0.8cm]
\includegraphics[width=.85\linewidth]{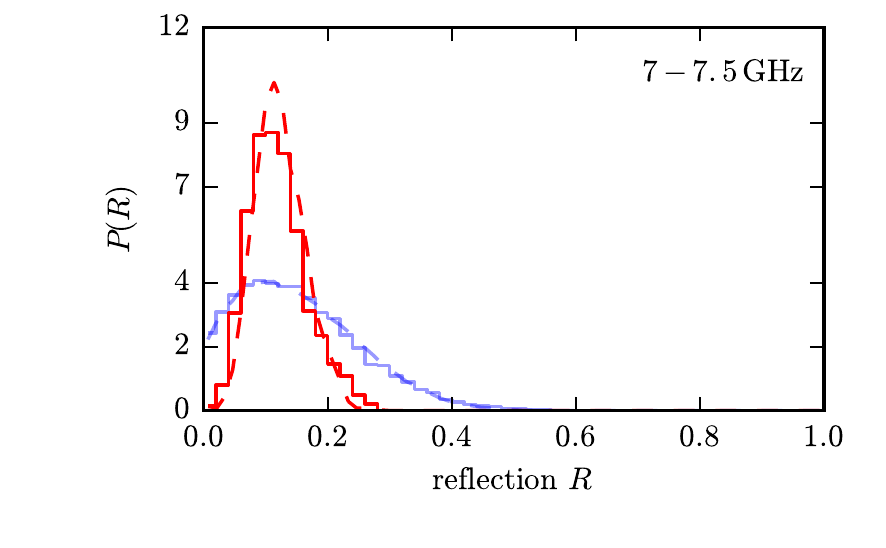}\\[-0.8cm]
\includegraphics[width=.85\linewidth]{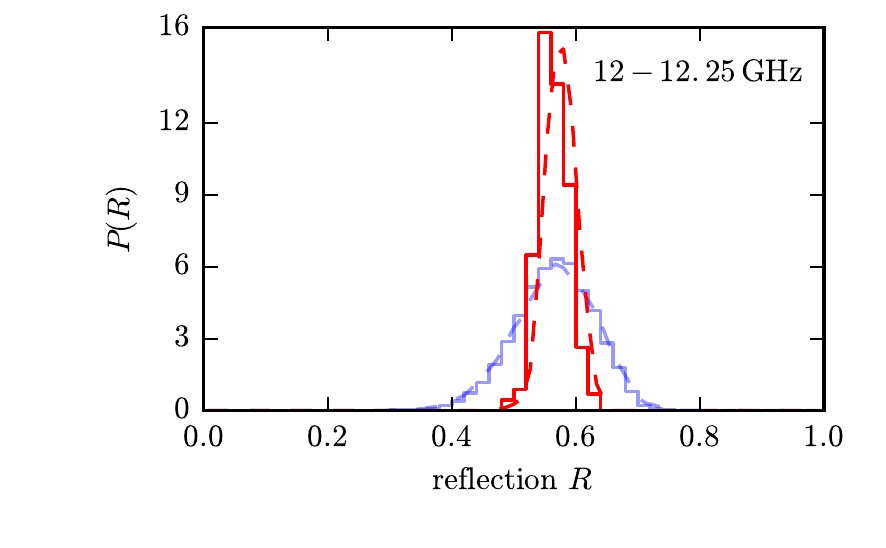}\\[-0.7cm]
\caption{\label{fig::P(R)}
Distribution of reflections $R$ for three different frequency ranges (top: 3-3.5, center: 7-7.5, bottom: 12-12.25 GHz), i.e. antenna coupling values,
and for two different different openings ($h_o$=0\,mm in blue and 24\,mm in green). For corresponding values refer to table~\ref{tab:InfoMeas}.
The histogram shows the experimental distribution and the dashed line the distribution obtained by RMT simulations (see text) using the experimental extracted values of antenna couplings $T_a$ and absorption $\gamma$.
}
\end{figure}

\begin{figure}[!t]
\centering
\includegraphics[width=.85\linewidth]{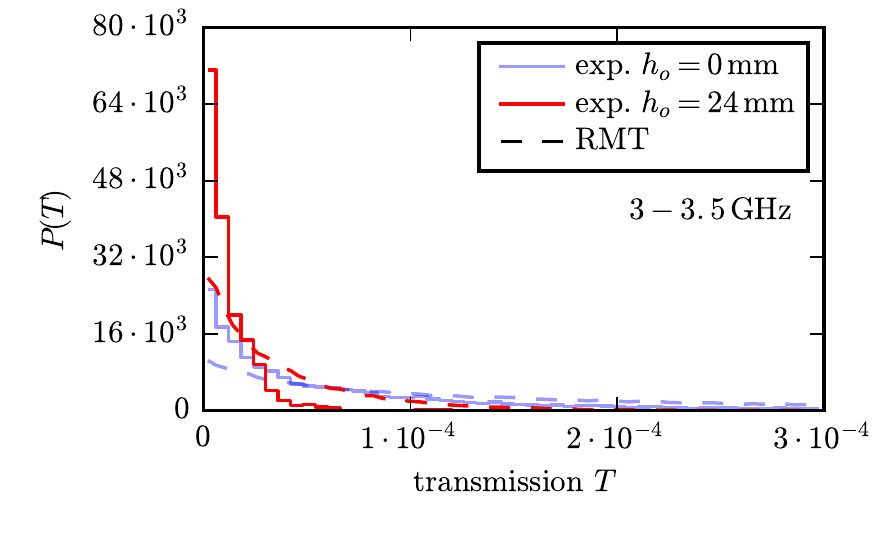}\\[-0.8cm]
\includegraphics[width=.85\linewidth]{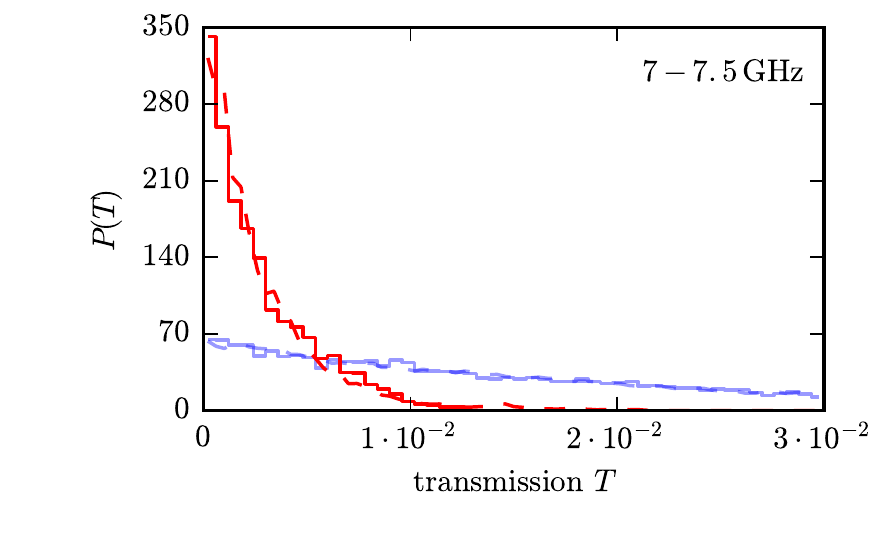}\\[-0.8cm]
\includegraphics[width=.85\linewidth]{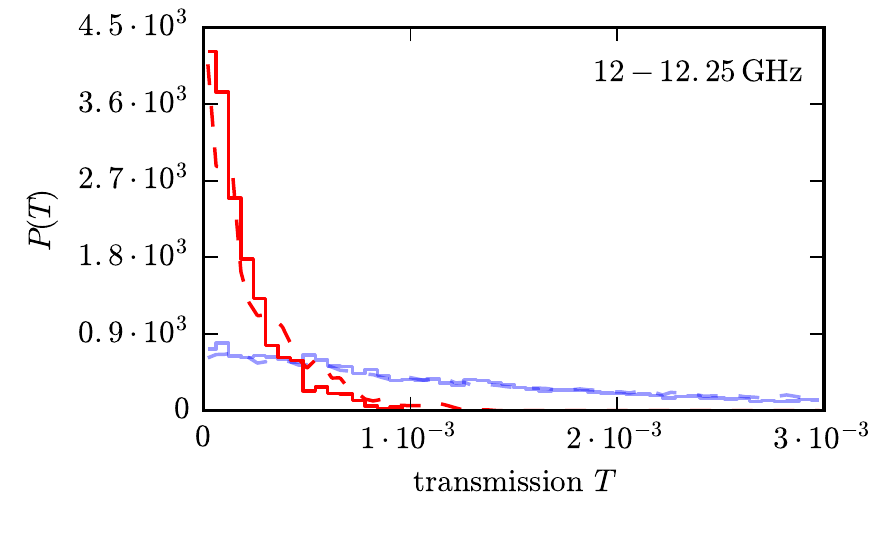}\\[-0.7cm]
\caption{\label{fig::P(T)}
As fig.~\ref{fig::P(R)} but for the distribution of the transmission $T$. Note the different scale of the abscissa.
}
\end{figure}

In fig.~\ref{fig::P(R)} and fig.~\ref{fig::P(T)} the experimental distribution of the reflection $R$ and transmission $T$ are presented as histograms for the three different frequency and the numerically obtained distributions. In general, we observe the behavior predicted by RMT though deviations are visible especially for larger openings.

\subsection{Average Spacing of Transmission Maxima}

\begin{figure}[!t]
\centering
\includegraphics[width=.9\linewidth]{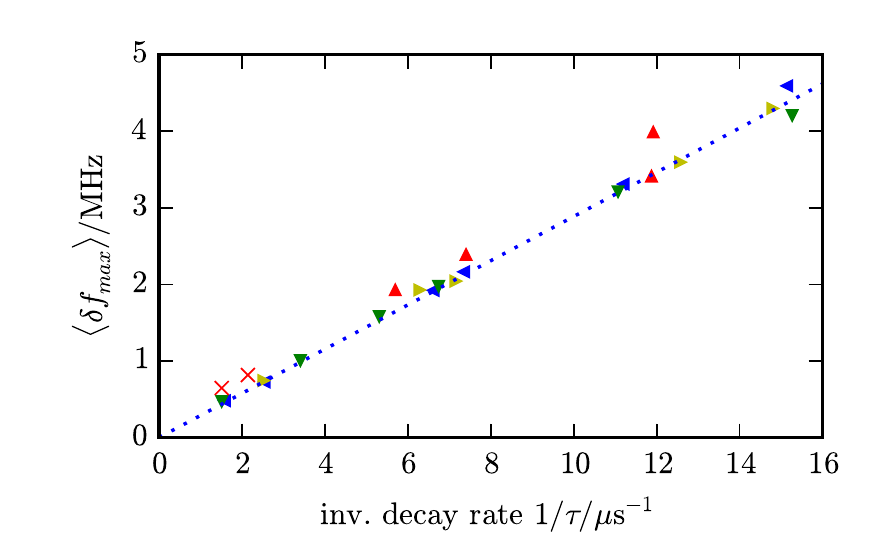}
\caption{\label{fig:dfmax}
Averaged frequency spacing of maxima $\langle \delta f_{max} \rangle$ in the transmission $T$ as a function of the inverse decay rate $\tau^{-1}$.
The upper red, left blue, right yellow, and green lower triangles correspond to the 4 different frequency regimes (3-3.5, 7-7.5, 7.5-8, and 12-12.25\,GHz), respectively.
The dotted blue line corresponds to the prediction from Schroeder and Kuttruff (see eq.~\ref{eq:df_max}) in the strong overlap regime\cite{sch62a}.
The red crosses mark the two results where the modal overlap $d$ is smaller than 3 (for 3-3.5\,GHz regime).
}
\end{figure}

Schroeder and Kuttruff investigated the behavior of the average spacing between maxima in the transmission $\langle\delta f_{max}\rangle$ of acoustic waves in rooms\cite{sch62a,kut09}.
They found that in the strongly overlapping regime, i.e. $d>3$ the relation between the $\langle\delta f_{max}\rangle$ and the intensity decay rate $\tau$ is given by
\begin{equation}\label{eq:df_max}
\langle\delta f_{max}\rangle=(2\sqrt{3}\tau)^{-1}.
\end{equation}
As most of the results obtained in this regime do not depend of the specific nature of the waves, we expect this relation is also valid in electromagnetic reverberation chambers.
In fig.~\ref{fig:dfmax} the spacing is plotted versus the inverse decay rate.
We find a good agreement between the theoretical prediction given by the blue dotted line and the different frequency regimes (see different triangles). For the low lying frequency regime most results are above the line, but even the values with a modal overlap close to $1$ (see red crosses in fig.~\ref{fig:dfmax}) are not far off.
This result provides an easy alternative estimate of the quality factor in a frequency band by just counting maxima rather than performing a cumbersome Fourier transform over a limited bandwidth.

\section{Conclusion}

We have shown that also for the three dimensional chaotic cavity the effective Hamiltonian in the strongly overlapping regime corresponds to the RMT statistics.
For large overlaps the predictions from the Ericson regime hold. Thus, the fields can be assumed to be isotropically distributed.
We verified the relation of the mean spacing of transmission maxima over a wide range of modal overlaps and quality factors.

\section*{Acknowledgment}

We would like to thank D. Savin for insightful discussions and the European Commission for financial support through
the H2020 programme by the Open Future Emerging Technology "NEMF21" Project (664828).

\end{document}